\documentclass[
superscriptaddress,
amsmath,amssymb,
aps,
prb,
twocolumn
]{revtex4-2}
\usepackage{graphicx}% Include figure files
\usepackage{dcolumn}% Align table columns on decimal point

\usepackage{bm}% bold math
\usepackage{xcolor}
\usepackage{bm}
\usepackage{siunitx}
\usepackage{booktabs}
\usepackage[colorlinks=true, citecolor=blue, linkcolor=blue]{hyperref}
\usepackage{changes}

\newcommand{\NS}{NiS$_{2}$}

\begin{document}

\preprint{APS/123-QED}

\title{Impact of strong electronic correlations on altermagnets: the case of \NS}

\author{Ina Park}
\affiliation{Center for Computational Quantum Physics (CCQ), Flatiron Institute, New York, New York 10010, United States}

\author{Turan Birol}
\affiliation{Department of Chemical Engineering and Materials Science, University of Minnesota, Minneapolis, Minnesota 55455, United States}

\author{Antoine Georges}
\affiliation{Coll\`ege de France, 11 Place Marcelin Berthelot, 75005 Paris, France}
\affiliation{Center for Computational Quantum Physics (CCQ), Flatiron Institute, New York, New York 10010, United States}
\affiliation{CPHT, CNRS, \'Ecole Polytechnique, IP Paris, F-91128 Palaiseau, France}
\affiliation{DQMP, University of Geneva, 24 quai Ernest-Ansermet, 1211 Geneva, Switzerland}

\author{Rafael M. Fernandes}
\affiliation{Department of Physics, The Grainger College of Engineering, University of Illinois Urbana-Champaign, Urbana, Illinois 61801, United States}
\affiliation{Anthony J. Leggett Institute for Condensed Matter Theory, The Grainger College of Engineering, University of Illinois Urbana-Champaign, Urbana, Illinois 61801, United States}

\date{\today}

\begin{abstract}

One of the distinguishing features of an altermagnet is that its spin-up and spin-down bands display a nodal momentum-dependent splitting even in the absence of spin-orbit coupling. While this property has been investigated in many weakly-correlated altermagnetic materials, the impact of strong electron-electron interactions on the spin-dependent electronic structure has remained little explored, particularly in metals. Here, we propose \NS\ as a prototypical strongly correlated metallic altermagnet. While at ambient pressure this compound is an altermagnetic Mott insulator, it undergoes a pressure-driven metal–insulator transition (MIT) while maintaining its altermagnetic ordered phase. By systematically comparing DFT, DFT+$U$, and DFT+DMFT calculations on the metallic altermagnetic phase near the MIT, we disentangle how strong static and dynamic correlations modify the electronic structure. Specifically, the spin splitting of the bands is modified not only through the enhancement of the local magnetic moment caused by static correlations, but also by the momentum-dependent bandwidth renormalization caused by dynamic correlations. Moreover, dynamic electronic correlations cause a pronounced lifetime asymmetry between the spin-up and spin-down quasiparticles, an effect that is amplified by the particle–hole asymmetry promoted by Hund’s correlations. Our results not only shed light on the rich landscape of correlation effects in metallic altermagnets, but also establishes \NS\ as a platform to investigate the interplay between Mott and Hund physics and altermagnetic order.

\end{abstract}

\maketitle

\section{\label{sec:intro}Introduction}

Altermagnetic (AM) order has recently been proposed as a new class of collinear magnetism that is distinct, on symmetry grounds, from both ferromagnetic (FM) and antiferromagnetic (AFM) orders\cite{vsmejkal2022beyond,vsmejkal2022emerging}. Interestingly, it has the intriguing characteristics of both a vanishing net magnetization, like AFM phases, and a splitting between the spin-up and spin-down bands without spin-orbit coupling, like FM phases \cite{Jungwirth2025symmetry}. 
These properties stem from the unique symmetries of an AM state, in which the opposite-spin sublattices of the crystal are related neither by a translation nor by inversion, but by a rotation that can be proper, improper, or part of a nonsymmorphic glide or screw operation \cite{Smejkal2020_AM}. Under these conditions, the magnetic moments are compensated macroscopically while the Kramers degeneracy is lifted in the absence of spin-orbit coupling (SOC), giving rise to a momentum-dependent spin-splitting of the bands with $d$-wave, $g$-wave, or $i$-wave symmetry. These features have ignited a flurry of research on these systems both due to their fundamental properties \cite{vsmejkal2022emerging,Bhowal2024,Schiff2024,Fernandes2024_AM,Cano2024,Agterberg2024,Attias2024,Thomale2025,Antonenko2025} and their possible use in technological applications \cite{Jungwirth2025spintronics,Bai2024spintronics}.

Thus, on general grounds, AM order is a consequence of two different types of interactions \cite{Jungwirth2024supefluid}: crystalline interactions, which enforce the local symmetry of the magnetic sublattices, and electronic interactions, which promote a compensated magnetic state. Given this broad mechanism, it is not surprising that density-functional theory calculations and symmetry analyses have found AM order to be realized, or proposed to be realized, in a large number of materials  \cite{mazin2021prediction,vsmejkal2022beyond,Facio2023,Gao2023,Mazin2023_FeSe,Jaeschke2024,Haule2025,Sodequist2024,Smolyanyuk2024,che2024realizing,che2025bilayer,Bhattarai2025,Gu2025,Duan2025,Smejkal2024_multiferroics}, including metals like CrSb \cite{Reimers2023,Li2024,Ding2024,Yang2024,Lusignature2025} and KV$_2$O$_2$Se \cite{Jiang2024discovery,Zhang2024crystal}, semiconductors like MnTe \cite{Mazin2023,Lee2023,Osumi2023,Krempasky2024,Amin2024}, and Mott insulators like La$_2$Mn$_2$Se$_2$O$_3$ \cite{Ji2025,Garcia2025,Mazin2025}, organic charge transfer salts~\cite{Valenti2024,iguchi2025magneto}, and ABO$_3$ orthorhombic perovskites \cite{Fernandes2024_AM,Bernardini2025,Naka2025} . The latter raises the important question of how strong electron-electron interactions impact the altermagnetic state, a topic that has received more attention recently \cite{Valenti2024,Sobral2024,Giuli2025,Ouyang202}. 

A particularly interesting regime is that of a strongly correlated metallic phase, realized for instance in a material close to an interaction-driven metal-insulator transition (MIT). In this regime, although quasiparticles can still be formally defined, they generally have small spectral weights and shorter lifetimes, which strongly affect the transport and thermodynamic properties of the material \cite{georges1996dynamical,imada1998metal,georges2024hund}. Most works on metallic altermagnets, however, have focused on weakly-correlated systems, where long-lived sharp quasiparticles are present. While recent theoretical studies have shown important effects of strong correlations on altermagnets \cite{Valenti2024,Sobral2024,Giuli2025,Ouyang202}, a systematic study of the different mechanisms by which correlations impact the altermagnetic electronic structure is desirable.

To shed light on this issue, here we focus on \NS. At ambient pressure, this pyrite material is a Mott insulator that undergoes a transition as lowering the temperature to a magnetically ordered state that preserves all crystalline symmetries of the paramagnetic phase \cite{Nishihara1975,Kikuchi1978,Matsuura2003,Yano2016}. As a result, this magnetic phase can be classified as an AM state \cite{yu2024spin}. Upon application of pressure, be it physical or chemical, \NS\ undergoes a MIT transition. Interestingly, even in the metallic side of the phase diagram, the compound still undergoes a magnetic transition towards the same magnetically ordered state that it displays at ambient conditions, as shown schematically in Fig.~\ref{fig:1}(a)  \cite{Gautier1975,Ogawa1979,imada1998metal, matsuura2000magnetic,Miyasaka2000,Wilson2008, perucchi2009pressure, friedemann2016large}. Therefore,  \NS\ provides an ideal framework to investigate the electronic properties of a strongly-correlated metallic altermagnet. 

In this paper, we systematically analyze the altermagnetic electronic structure of \NS\ in the correlated metallic phase near the pressure-induced MIT. To disentangle the effects of static and dynamic correlations, we perform a comparative analysis using DFT (density-functional theory), DFT+$U$, and DFT+DMFT (dynamical mean-field theory). The key difference between the latter two are how interactions are treated: while DFT+$U$ treats the interactions in a static mean-field approach, DFT+DMFT treat them dynamically by solving an auxiliary impurity problem. \cite{anisimov1997first, georges1996dynamical, carta2025explicit} This is a crucial difference, since the quasiparticle lifetime is determined by the frequency dependence of the self-energy, which is absent in DFT+$U$. The crucial impact of dynamic correlations on the quasiparticle lifetime in \NS\ was indeed observed in previous DMFT studies, which however focused only on the paramagnetic phase \cite{Matsuura1998,jang2021direct,DayRoberts2023,park2024clean}.

Here, by focusing on the altermagnetic phase, we find that interactions generally increase the local magnetic moment on the Ni $d$ orbitals, which can enhance the spin splitting between the spin-up and spin-down bands. In DFT+$U$, this moment-enhancement-driven appears relatively uniform because the correlation are static. In contrast, in DFT+DMFT, the correlation-induced modification of the spin splitting is the result of two competing effects. The increased local moment tends to enhance the splitting whereas the bandwidth renormalization due to dynamic correlations tends to suppress it, resulting in a strongly momentum- and energy-dependent spin-splitting modification.
% Here, by focusing on the altermagnetic phase, we find that interactions generally increase the local magnetic moment on the Ni-$d$ orbitals which, in turn, enhances the splitting between the spin-up and spin-down bands of the metal. While in the DFT+U this enhancement is essentially uniform across all bands, due to the static character of the correlations, in DFT+DMFT the changes in the spin splitting depend strongly on the bands' spectral weight for the particular momentum and energy values.
Interestingly, we observe that the sulfur-derived bands, which are nominally nonmagnetic, also acquire significant spin splitting due to the hybridization with the Ni-$d$ states. Besides the band-dependent changes in the spin-splitting, we find that dynamic correlations cause a pronounced anisotropy in the lifetimes of the spin-up and spin-down quasiparticles, which is roughly proportional to the splitting between the bands. This can have important consequences for spintronics applications that manipulate carriers with a specific spin quantum number.  Overall, our results highlight the qualitative effects of strong correlations on the altermagnetic electronic structure, and establish \NS\ as a representative platform to explore the interplay between correlation effects and AM order.

Our paper is organized as follows: in Sec.\ref{sec:r1}, we discuss the crystal structure and the altermagnetic phase of \NS\, analyzed in detail based on group theory and DFT calculation. Then in Sec.\ref{sec:r2}, we discuss the altermagnetic electronic structure obtained from the DFT+DMFT calculation, and in Sec.\ref{sec:r3}, we systematically compare DFT, DFT+$U$, and DFT+DMFT electronic structures and investigate how static and dynamic correlation effects can modify the spin splitting sizes of the bands. Finally, in Sec.\ref{sec:r4}, we discuss the correlation effects on the spin-resolved quasiparticle lifetime focusing on its pronounced anisotropy between the different spin bands originating from dynamic correlation effects.

\section{\label{sec:r1}Crystal structure and altermagnetic phase of \NS}

\begin{figure}
    \centering
    \includegraphics[width=0.99\linewidth]{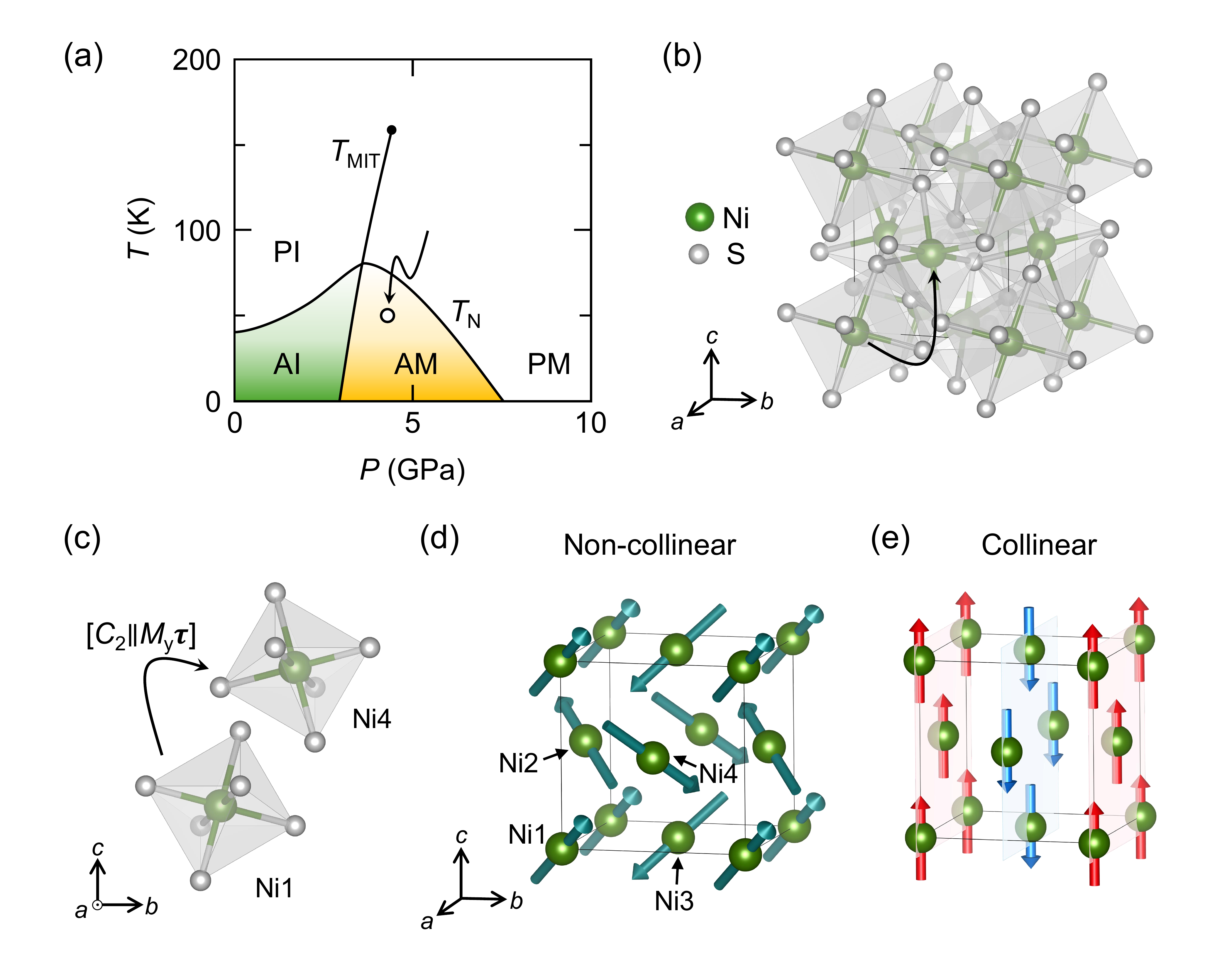}
    \caption{(a) Schematic pressure ($P$) - temperature ($T$)  phase diagram of \NS. \cite{friedemann2016large} Open circle indicates the $P,\,T$ values at which the DMFT calculations are performed. (b) Crystal structure of the pyrite \NS. (c)  Example of a symmetry relation between two representative Ni atoms with opposite spins The symbol $\left[ C_2 \parallel M_y t \right]$ indicates that the system is invariant under a combination of a two-fold rotation in spin space and a non-symmorphic glide $M_y t$ in real space, with $t= (0, 1/2, 1/2)$ (d) The non-collinear magnetic structure realized experimentally in \NS\ and (e) the corresponding approximate collinear structure. }
    \label{fig:1}
\end{figure}

The crystal structure of \NS\ is shown in Fig.\ref{fig:1}(b). It is a pyrite structure with a simple cubic unit cell, where four Ni atoms located at face-center sites are surrounded by sulfur (S$_6$) octahedra. 
These octahedra, which generate anisotropic crystal fields on the Ni atoms, are related to each other by $C_3$ rotations, as well as nonsymmorphic glide operations. These ``$a$'' glide operations in the space group involve a mirror reflection on a plane in the $\{001\}$ family, followed by a translation by half a lattice vector on the same plane. 
%
%These octahedra, which generate an anisotropic electrostatic potential on the Ni atom, are related by a nonsymmorphic mirror symmetry \rmf{Ina, I'm a bit confused by what is meant with nonsymmorphic mirror symmetry. It does not seem to be a glide, since the translation has a component perpendicular to the mirror. It is not a screw axis either. What is it?}. \ina{I used 'nonsymmorphic' because of fractional translation, I understood it as we need mirror $M$ and translation $t$=(0,/1,2/1,2) so it's not symmorphic. Please correct this if I'm wrong.} 
%
Indeed, as shown in the figure, the octahedra are tilted with respect to the lattice vectors and are not parallel to each other. In particular, each of the four octahedra in the unit cell is slightly distorted, and has a different unique axis oriented along one of the four $\langle 111 \rangle$ axes. As a result, they cannot be related by a simple translation. For example, Ni1 and Ni4, the atoms connected with an arrow in Fig.\ref{fig:1}(b), can be mapped onto each other via $M_yt$, where $M_y$ is a mirror with respect to the plane perpendicular to the $b$ axis and $t$ is the half translation $t = (0, 1/2, 1/2)$, as depicted in Fig. \ref{fig:1}(c).

Experimentally, the magnetic state of \NS\ that onsets at the higher temperature (39~K) is known to be a non-collinear compensated magnetic state  \cite{Nishihara1975,Kikuchi1978,Matsuura2003,Yano2016}.  It is noteworthy that this phase has a magnetic ordering wave-vector of $\vec{q} = (0, 0, 0)$ and therefore does not require a supercell construction. The magnetic order parameter transforms as the $A_{1g}^-$ irreducible representation of the parent point group, and breaks only the time-reversal symmetry. As shown in Fig.\ref{fig:1}(d), in this non-collinear phase, the local magnetic moments at the four Ni atoms are aligned in a compensated manner along the local $\langle 111\rangle$ directions corresponding to the unique $C_3$ rotation axis of the local site symmetry (point) group. This type of order in which the magnetic sublattices are related by a non-symmorphic mirror corresponds to an AM state (see also Ref. \cite{yu2024spin}). The non-collinearity of the spins in this case is a result of the spin-orbit coupling (SOC) that aligns the antiparallel moments along the local $\langle 111 \rangle$ axes. 

As shown in Fig.\ref{fig:1}(a), this AM phase persists in a wide range of pressure in the phase diagram of \NS, from ambient pressure to about $P=7$ GPa \cite{friedemann2016large}. This is because the cubic lattice symmetry is preserved throughout the pressure range, with a monotonic change of the lattice parameter both in the insulating and metallic regions. Between the two regions, the MIT first-order transition line crosses the dome of magnetic order. Therefore, \NS\ exhibits a unique transition from an AM insulator to an AM metal, with the MIT preserving the magnetic long range order.  

To further elucidate the AM character of this phase, we use the classification scheme of Ref. \cite{Fernandes2024_AM} for AM in the presence of SOC. The space group of  \NS\ is Pa$\bar{3}$ ($\#205$), with point group $T_h$. Since the non-collinear magnetic configuration in Fig.\ref{fig:1}(d) preserves all crystalline symmetries, the magnetic order parameter transforms as the irreducible representation (irrep) $A_{g}^-$ of the point group (or, equivalently, irrep $m\Gamma_1^+$ of the space group). Here, to indicate that an irrep is odd under time reversal, we use a minus superscript for Mulliken symbols and an $m$ pre-factor for irreps of the space group. Condensation of  $\Phi_{\mathrm{ncl}}$ leads to the cubic magnetic space group Pa$\bar{3}.1$ ($\#205.33$). Following  Ref. \cite{Fernandes2024_AM}, this irrep corresponds to a ``pure'' AM order parameter. Its effect on the low-energy electronic degrees of freedom can be described in terms of an effective Hamiltonian 

\begin{equation}
    H_{\mathrm{AM}}^{\mathrm{ncl}} = \sum_{\mathbf{k}}\Phi_{\mathrm{ncl}}\left( k_xk_y\sigma_z + k_yk_z\sigma_x+k_zk_x\sigma_y\right) \psi^{\dagger}_{\mathbf{k}} \psi^{\phantom{\dagger}}_{\mathbf{k}}\,,
    \label{eq:1}
\end{equation}

where $\psi_{\mathbf{k}}$ is a spinor and $\Phi_{\mathrm{ncl}}$ is a parameter proportional to the magnetic order parameter.  Eq. \ref{eq:1} corresponds to a momentum-dependent spin splitting and exhibits a $d$-wave spin texture, with nodal lines along the main axes $k_x=k_y=0$,  $k_x=k_z=0$, and  $k_y=k_z=0$. 

As explained above, without SOC, altermagnetism is a collinear compensated magnetic phase. It is therefore convenient to consider a collinear version of the magnetic order pattern of \NS\ that has a similar spin texture. Such a collinear phase is also much more amenable to be treated via DFT+DMFT. In group-theory language, we seek a collinear phase that, without SOC, is invariant under $\left[ C_2 \parallel M_y t \right]$, where the left operation $C_2$ corresponds to a two-fold rotation of the spins with respect to a transverse axis and the right operation acting on the crystal is the non-symmorphic mirror $M_y t$. We find that such a collinear phase consists of two up-spin (``up'') and two down-spin (``dn'') distributed on the Ni atoms as shown in Fig. \ref{fig:1}(e), corresponding to parallel spins along the $ac$ plane and anti-parallel spins along the $b$ axis. Interestingly, this collinear phase has often been employed in first-principles calculations without SOC as an approximate form of the non-collinear phase, allowing one to avoid computational complexity. \cite{schuster2012electronic}

To show that this collinear phase gives a similar spin splitting as the non-collinear phase, we turn on the SOC again to treat them on an equal footing. In the collinear phase with SOC, the magnetic order parameter transforms as the irreducible representation (irrep) $(E_{g}^{*})^-$ of the point group (or the irrep $m\Gamma_2^+\,m\Gamma_3^+$ of the space group), such that its condensation leads to the orthorhombic magnetic space group Pbca$.1$ ($\#61.433$). The fact that the crystal symmetry in the magnetic phase is lowered to orthorhombic follows in a straightforward way from the magnetic configuration shown in Fig. \ref{fig:1}(e). In contrast, the reason why the non-collinear configuration preserves the cubic point group is because the spins point along high-symmetry axes of the Ni-S tetrahedra. Using group theory, we find that the manifestation of the collinear order on the low-energy electronic degrees of freedom can be described in terms of the effective Hamiltonian 

\begin{equation}
 H_{\mathrm{AM}}^{\mathrm{cl}} =     \sum_{\mathbf{k}}\Phi_{\mathrm{cl}}\left( A_z k_xk_y\sigma_z + A_x k_yk_z\sigma_x+A_y k_zk_x\sigma_y\right) \psi^{\dagger}_{\mathbf{k}} \psi^{\phantom{\dagger}}_{\mathbf{k}}\ \, ,
    \label{eq:2}
\end{equation}

 where $\Phi_{\mathrm{cl}}$ is a parameter proportional to the magnetic order parameter. Eq.~\ref{eq:2} has a similar functional form as $H_{\mathrm{AM}}^{\mathrm{ncl}}$, in that the spin-split nodal lines are located along the main axes. The main difference is that the coefficients $A_i$ are no longer constrained to be equal, since the cubic symmetry is reduced to an orthorhombic symmetry in the collinear phase. Thus, the collinear phase can be interpreted as a symmetry-reduced projection of the non-collinear magnetic phase, for example, onto the $z$ component.

\begin{figure}
    \centering
    \includegraphics[width=0.95\linewidth]{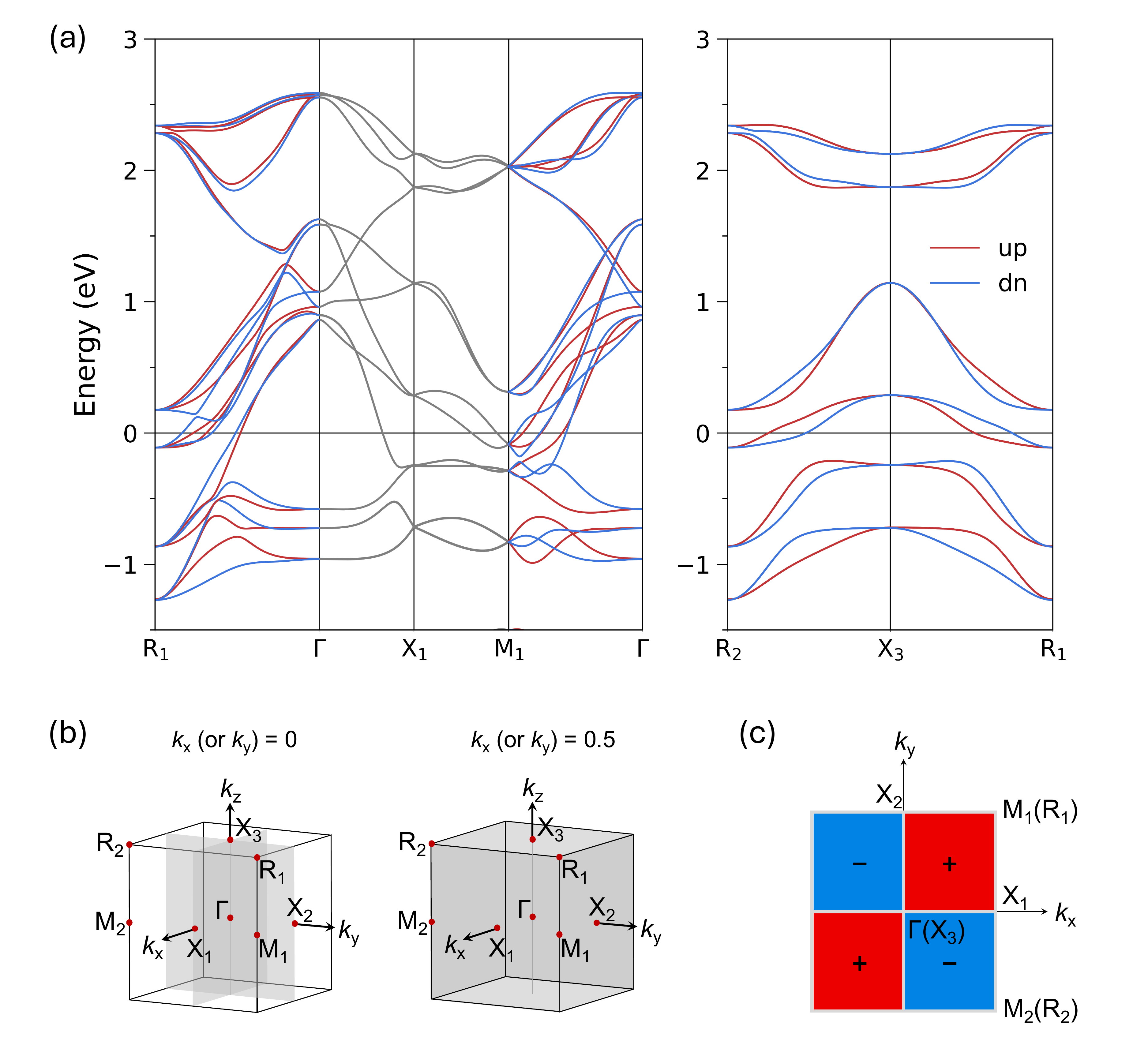}
    \caption{(a) Non-relativistic DFT band structure for the collinear altermagnetic phase of metallic \NS\ at $P=4.3$ GPa. Grey lines are spin-degenerate band dispersions. (b-c) Brillouin zone and high-symmetry $k$-points (b) with spin-degenerate nodal planes and (c) with non-zero spin-splitting at the $k_z=0$ ($k_z=1/2$, $k$-points in parenthesis) planes. Red and blue colors highlight the alternating sign of the splitting between spin-up and spin-down bands.}
    \label{fig:2}
\end{figure} 

This group theory analysis is in agreement with our non-relativistic spin-polarized DFT calculations for the collinear magnetic phase (see Appendix.\ref{app:b} for details). The DFT calculations were performed at $P$ = 4.3 GPa, which is close to the MIT (open circle in Fig.~\ref{fig:1}(a)). As shown in Fig.~\ref{fig:2}(a)-(b), the spin-polarized band structure displays nodal planes at $k_x,\,k_y=0$ and at the zone boundaries $k_x,\,k_y=1/2$, as expected from Eq. (\ref{eq:2}) if SOC is turned off. Moreover, these figures show a sizable spin splitting, away from the nodal planes, even though SOC is not included in the calculation.
The $d$-wave symmetry of the spin splitting can be directly seen in the right panel of Fig.~\ref{fig:2}(a), as the splitting between spin-up and spin-down bands is reversed between the two directions that are rotated by $90^\circ$. This region of non-zero spin splitting is schematically depicted in Fig.~\ref{fig:2}(c), highlighting its $d$-wave character.

\section{\label{sec:r2} Altermagnetic electronic structure near the MIT}

\begin{figure*}
    \centering
    \includegraphics[width=1.0\linewidth]{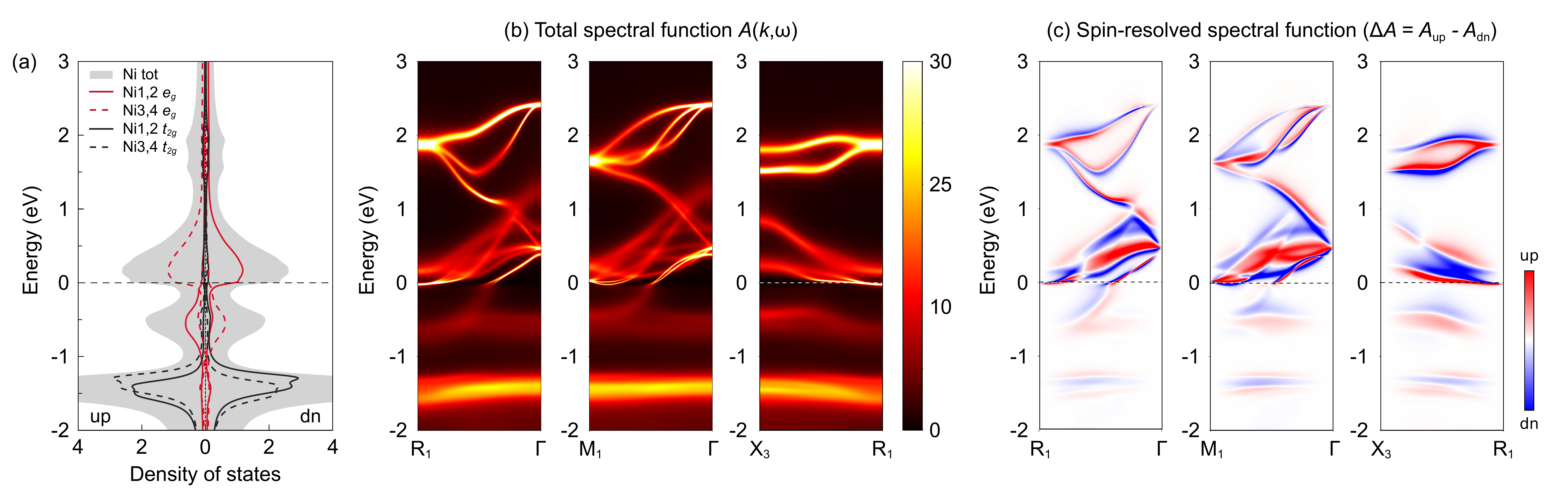}
    \caption{(a) Projected density of states (PDOS) of Ni $e_g$ and $t_{2g}$ orbitals. The labels are the same as in Fig. \ref{fig:1}(d).  (b) Total momentum-resolved spectral function $A(k,\omega)$ and (c) spin-resolved spectral function defined as $\Delta A \equiv A_{\mathrm{up}}(k,\omega) - A_{\mathrm{dn}}(k,\omega)$ along the momentum-space paths R$_1$-$\Gamma$ (left column), M$_1$-$\Gamma$ (middle), and X$_3$-R$_1$ (right). The Brillouin zone notation is the same as in Fig. \ref{fig:2}(b).}
    \label{fig:3}
\end{figure*}

As discussed above, \NS\ undergoes a MIT within the AM dome. By virtue of this unique phase diagram, \NS\ can serve as a prototypical example to investigate the effects of tunable strong electronic correlations on the altermagnetic electronic structure. Thus, here we study the correlated metallic phase near the MIT (at $P$ = 4.3 GPa and $T$ = 50 K), as indicated by the open symbol in Fig. \ref{fig:1}(a), focusing on the changes in both the altermagnetic splitting size and on the spin-polarized quasiparticle lifetimes. To capture both effects, we employ the density functional theory + embedded dynamical mean-field theory (DFT+DMFT) method~\cite{georges1996dynamical, haule2010dynamical, haule2015free}. DFT+DMFT is a sophisticated method which can describe many body correlation effects in real materials beyond DFT+$U$ \cite{anisimov1997first, lichtenstein1998ab,kotliar2006electronic}. The difference between these two approaches is further discussed below.

The details of the DFT+DMFT calculation methods are described in Appendix~\ref{app:b}, and the results are shown in Fig.~\ref{fig:3}. Fig.~\ref{fig:3}(a) displays the density of states (DOS). We can see that while the total DOS of the Ni atoms with up (labeled ``up'') and down (labeled ``dn'') spin components are exactly compensated, the local spin-resolved projected DOS (PDOS) on the four Ni atoms split into two magnetic sublattices -- i.e., the PDOS of Ni1 and Ni2 are the same (black and red solid lines) and the PDOS of Ni3 and Ni4 are the same (black and red dashed lines). The atoms labeled Ni1(2) and Ni4(3) in Fig. \ref{fig:1}(d) have ``up'' and ``dn'' majority spins, respectively. As a result, the spectral weight distribution are the same between the different atoms and spin components, e.g. between Ni1-up and Ni4-dn, so that the total ``up'' and ``dn'' DOSs become equal, resulting in a vanishing net magnetization.

The PDOS also shows that the main contribution to the magnetic moment comes from the Ni-$e_g$ orbitals, which have a ground-state spin $S = 1$ due to the octahedral ligand field and the nominal occupancy of Ni$^{2+}$. In our DFT+DMFT calculation, the resulting magnetic moment size was obtained as M$_{\mathrm{DMFT}} = 1.11 $ ($\mu_\mathrm{B}/\mathrm{atom}$) at $T=50$ K, which is consistent with the experimental value of M$_{\mathrm{exp}} = 1.15$ $(\mu_\mathrm{B}/\mathrm{atom}$) measured using neutron diffraction at $T=10$ K ~\cite{matsuura2000magnetic}.

Figs.~\ref{fig:3}(b) and ~\ref{fig:3}(c) show the total and spin-resolved band structure, which are defined as $A(\mathbf{k},\omega)=A_{\mathrm{up}}+A_{\mathrm{dn}}$ and $\Delta A(\mathbf{k},\omega)=A_{\mathrm{up}}-A_{\mathrm{dn}}$, respectively. The spin-resolved band structure shows a spin-splitting symmetry that is consistent with the non-relativistic DFT results shown in Fig. \ref{fig:2}(a). The effective mass was also extracted from the self-energy of the majority spin components (see Fig.S~\ref{fig:s1}(b)), which dominantly contributes to the so-called $\alpha$-band \cite{xu2014direct}, which is the band enclosing the $\Gamma$ point in Fig.~\ref{fig:3}(b) and responsible for the large hole pocket at the Fermi surface. The obtained value is $m_{\mathrm{eff}}^{\mathrm{DMFT}} \sim 6.25$ and consistent with the experimental value of $m_{\mathrm{eff}}^{\mathrm{exp}} \sim 5.8$ ~\cite{semeniuk2023truncated}, measured via quantum oscillations. 

\section{\label{sec:r3}Correlation effects on the altermagnetic spin-splitting}

In this section, we systematically investigate how strong electronic correlations modify the amplitude of the AM spin splitting, defined for each band $\nu$ as $\Delta_{\mathrm{AM}}^{\nu}\left(\mathbf{k}\right)\equiv \varepsilon^{\nu}_{\mathrm{up}}\left(\mathbf{k}\right) - \varepsilon^{\nu}_{\mathrm{dn}}\left(\mathbf{k}\right)$. On general grounds, we expect that a larger interaction will lead to both an increase of the local magnetic moment and 
a suppression of the bandwidth (or an enhancement of the effective mass). Interestingly, while the larger moment should enhance the spin splitting, the smaller bandwidth is expected to reduce it.

To disentangle these two effects, we systematically compare the DFT, DFT+$U$, and DFT+DMFT band structures of metallic \NS\ under pressure. DFT+$U$ is essentially a \textit{static} mean-field treatment of the local electron-electron interaction, which enhances the static magnetic moment and produces orbital-dependent energy shifts that often resemble quasi-rigid band displacements \cite{anisimov1997first,carta2025explicit}. On the other hand, DFT+DMFT employs a \textit{dynamical} mean-field treatment of the interaction, thus capturing local many body effects manifested in the frequency dependence of the self-energy \cite{georges1996dynamical,georges2004strongly}. As a result, DFT+DMFT captures additional effects beyond DFT+$U$, such as bandwidth renormalization and finite quasiparticle lifetimes.

In order to properly compare the electronic structure obtained from DFT+DMFT to that of DFT and DFT+$U$, we will consider the \textit{quasiparticle} bands obtained by solving the quasiparticle equation:
\begin{equation}
    \mathrm{det}\left [(\omega + \mu - \varepsilon_{\mathbf{k},\nu})\delta_{\nu\nu'} - \mathrm{Re}\Sigma_{\nu\nu'}(\mathbf{k},\omega)\right] = 0 \, ,
    \label{eqn:3}
\end{equation}
in which $\nu$ is a band index, the imaginary part of the DMFT self-energy has been neglected, and the solutions $\omega=\omega_\nu(k)$ yield 
the quasiparticle excitations in band $\nu$. In this expression, the DMFT self-energy is ``upfolded'' from the orbital basis $\chi_m$ to the Bloch basis $\psi_{\mathbf{k}\nu}$: 
\begin{equation}
    \Sigma(\mathbf{k}, \omega)_{\nu\nu'} = \sum_{mm'} {\langle \psi_{\mathbf{k}\nu}| \chi_m\rangle \Sigma_{mm'}(\omega)\langle\chi_{m'}|\psi_{\mathbf{k}\nu'}\rangle }\,.
    \label{eq:4}
\end{equation}

\begin{figure*}[t]
    \centering
    \includegraphics[width=0.8\linewidth]{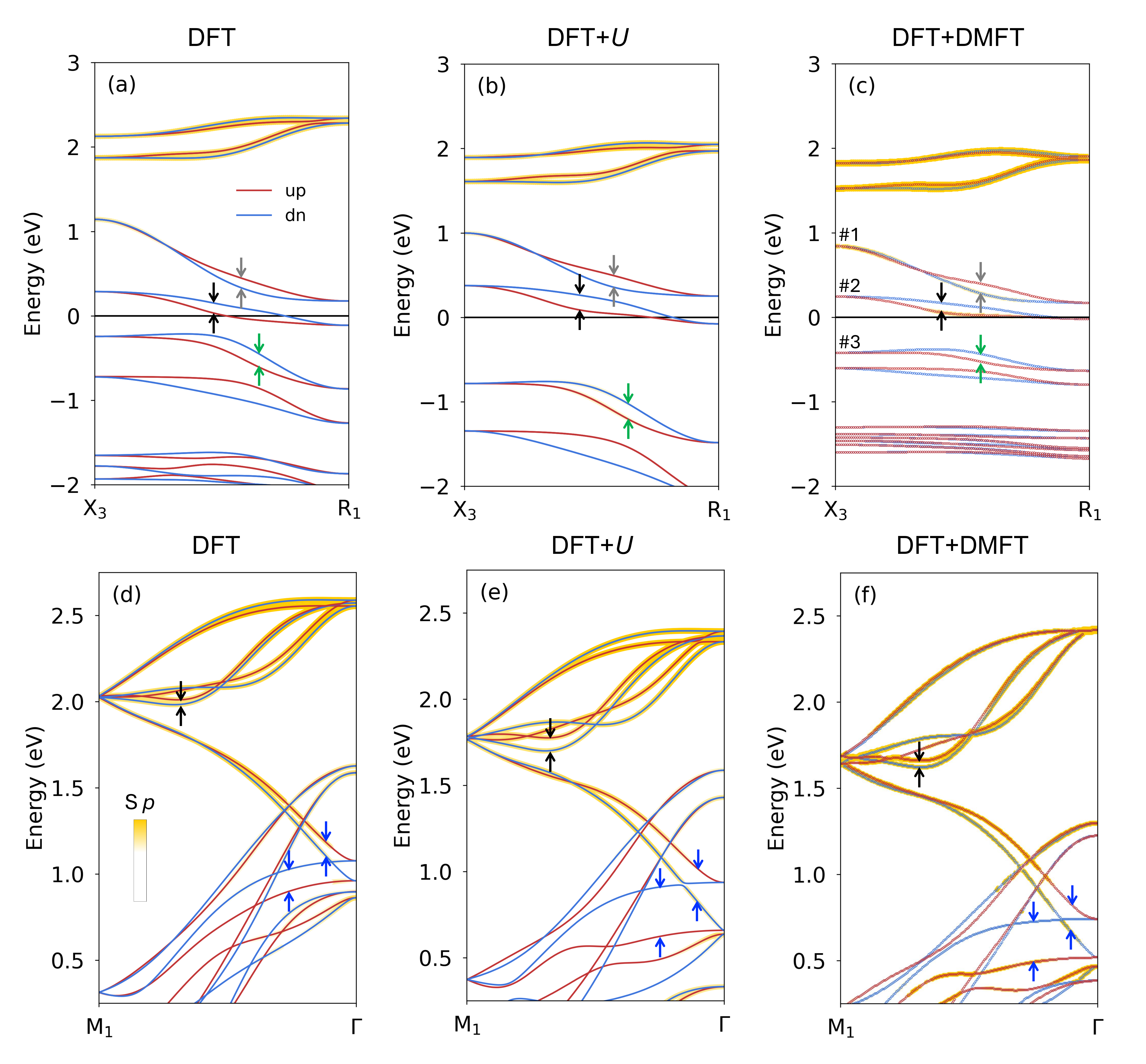}
    \caption{Comparison between (a)-(d) DFT, (b)-(e) DFT+$U$, and (c)-(f) DFT+DMFT effective band structures along the X$_3$-R$_1$ (upper panel) and M$_1$-$\Gamma$ (lower panel) paths.  The arrows highlight the AM spin-splitting of selected pairs of bands. The yellow shading gives the spectral weight of the S-$p$ orbitals.}
    \label{fig:4}
\end{figure*}

In practice, it is more convenient to determine the dispersions $\omega_\nu(\mathbf{k})$ by locating the maxima of the momentum (or energy) distribution curves associated 
with the spectral function after replacing the imaginary part of the self-energy by a small broadening parameter.  The exact procedure used in this work is explained in detail in Supplementary Sec. \ref{app:d}. 

Fig.~\ref{fig:4} compares the DFT, DFT+$U$, and DFT+DMFT band structures along two representative paths where the spin splitting is maximum: X$_3$-R$_1$ and M$_1$-$\Gamma$ (see Fig. \ref{fig:2}(b) for the Brillouin zone convention).  In this figure, the yellow shading in the fat bands refer to the S $p$-orbital weight. 
%
%In this figure, instead of $A(k,\omega)$ shown in Fig.~\ref{fig:3}, we show the 'effective' DFT+DMFT band structure to unambiguously compare the spin splitting size. The effective band structure was extracted from the spectral function with setting the imaginary part of the self-energy as zero, and further details are described in the Appendix~\ref{app:d}.
%
First, comparing DFT with DFT+$U$ shows that the the AM spin splitting is enhanced in DFT+$U$. As shown in Figs.~\ref{fig:4}(a)-(b) and Figs.~\ref{fig:4}(d)-(e), the splitting between each ``up'' and ``dn'' pair of closest bands increases once the interactions are introduced, as highlighted by the arrows. For the DFT+$U$ calculation, the parameter $U = $  2 eV was set to give a local magnetic moment of M$_{\mathrm{DFT}+U} = 1.10 $ ($\mu_\mathrm{B}/\mathrm{atom}$) that is consistent with the experimental and DFT+DMFT results (see Appendix ~\ref{app:b} for further calculation details). It's also noteworthy that DFT+$U$ does not only enhance the spin splitting but also shifts the correlated bands to higher binding energies. This tendency is clearly visible for the bands at around $-1$ eV to $0$ eV, where the static Hubbard term in DFT+$U$ starts to mimic the onset of Hubbard sub-bands in proximity to the MIT. 

While the DFT+$U$ results show a shift in the bands, this is essentially a rigid shift that has a negligible impact on the bandwidth. This is ultimately a consequence of the fact that only static correlations are present. To assess the impact of the bandwidth renormalization, we compare the band structures obtained from DFT+$U$ and DFT+DMFT in Figs.~\ref{fig:4}(b)-(c) and Figs.~\ref{fig:4}(e)-(f). Because the dynamical correlations described in DFT+DMFT renormalize the bands in a momentum- and energy-dependent manner, the overall bandwidth is renormalized according to Eq.\ref{eq:4} above. Moreover, because the $U$ value in the DFT+$U$ calculations was chosen to give the same magnetic moment magnitude as the DFT+DMFT calculations, this comparison allows us to disentangle the effects of enhanced moments and band renormalization.  
%\begin{equation}
%    \Sigma(k, \omega)_{\nu\nu'} = \sum_{mm'} {\langle \psi_{k\nu}| \chi_m\rangle \Sigma_{mm'}(\omega)\langle\chi_{m'}|\psi_{k\nu'}\rangle },
%    \label{eq:3}
%\end{equation}
%which will give DFT+DMFT band energies through: 
%\begin{equation}
%    G_{k, mm'}^{-1}(\omega) =  \omega + \mu - \varepsilon_{k,mm'} - \Sigma_{mm'}(\omega).
%    \label{eq:4}
%\end{equation}
Eq.\ref{eq:4} shows that the band renormalization comes from the real part of the local self-energy for each orbital, weighted by the 
matrix elements specifying the orbital composition of the band at each momentum. This effect is quantified by the momentum-dependent quasiparticle weight for a given band, which is given by 
$1/Z_\nu(\mathbf{k})= \sum_m |\langle \psi_{\mathbf{k}\nu}| \chi_m\rangle|^2/Z_m$, where $Z_m$ is the quasiparticle residue of the basis orbital $m$. 

While a more detailed analysis of the connection between the quasiparticle weight and the spin splitting magnitude is given in Appendix~\ref{app:c}, here we present the main results. We find that the correlation-induced change in the non-interacting value of the spin-splitting is not uniform. For instance, focusing on the bands near the Fermi level, some of the Ni-$e_g$ derived bands (highlighted by the pairs of arrows in Fig.~\ref{fig:4}(c)) show a spin splitting that is either unchanged or slightly enhanced  (black and gray arrows), while the other spin splitting of the other band shows a strong reduction (green arrows). This is in sharp contrast with the DFT+$U$ results, in which the  spin splitting of all bands are enhanced. This contrasting behavior highlights that once dynamical correlations are included, the changes in the spin splitting become strongly band- and energy-dependent, 
rather than an essentially uniform change. 
Such a ``redistribution'' of spin splitting is an orbital-weight dependent effect, which follows from the definition of the quasiparticle weight $1/Z_{\nu}$. Indeed, bands with stronger Ni-$e_g$ character are more heavily renormalized than those dominated by Ni-$t_{2g}$ or Se-$p$ orbitals, reflecting how the contribution of the self-energy is weighted by the orbital composition of each band. As a result, the modification of the AM spin splitting is governed not only by the overall strength of the correlations but also by the orbital composition of the state involved.

% \begin{figure*}[t]
%     \centering
%     \includegraphics[width=0.95\linewidth]{Figure 4_horizontal.pdf}
%     \caption{Comparison between (a)-(d) DFT, (b)-(e) DFT+$U$, and (c)-(f) DFT+DMFT effective band structures along the X$_3$-R$_1$ (upper panel) and M$_1$-$\Gamma$ (lower panel) paths.  The arrows highlight the AM spin-splitting of selected pairs of bands. The yellow shading gives the spectral weight of the S-$p$ orbitals. \rmf{What does the shading in panels (c) and (f) correspond to?} \ina{What do you mean by shading?} \rmf{the panels look small, is there another configuration that we can use to make them bigger?}}
%     \label{fig:4}
% \end{figure*}

In addition, the change in the spin splitting is further controlled by the intrinsic particle-hole asymmetry of the DMFT self-energy. Even when the orbital weights are comparable, bands on the hole side (negative energies) experience a stronger reduction of the splitting than those on the electron side (positive energies). This is a consequence of the fact that the real part of the self-energy maintains a nearly linear frequency dependence over a relatively broad range of negative frequencies, while at positive frequencies it quickly departs from linearity and becomes weakly frequency-dependent. The deviation from the quasiparticle linear behavior can be viewed as a ``kink'' feature of electronic origin (see Fig.S~\ref{fig:s1} and texts in the Supplementary section \ref{app:e}). 
As further discussed in the next section, the Hund's coupling 
is known to significantly contribute to the correlation effects, being related to the kink feature and to the
particle-hole asymmetry in \NS\ near the MIT 
(induced by both chemical substitution and pressure)~\cite{jang2021direct, park2024clean}. 
The change of the AM splitting is therefore not only orbital-dependent but also frequency-asymmetric, reflecting the Hund's correlation character of the system.  
Strong particle-hole asymmetry is indeed a hallmark of 
Hund metals~\cite{georges2013strong, georges2024hund}. Thus, we conclude that the outcome of the two competing correlation-driven effects on the spin splitting -- magnetic moment enhancement and bandwidth renormalization -- depends on the orbital composition of the band and whether the band is at positive or negative energies.

% \begin{figure*}[!htbp]
%     \centering
%     \includegraphics[width=0.90\linewidth]{Figure 5.pdf}
%     \caption{(a)-(d) DFT+DMFT band structures, with the corresponding quasiparticle lifetime marked by colored circles along the (a) $\Gamma$-R$_1$ and (b)-(d) X$_3$-R$_1$ paths. (e)-(h) Altermagnetic spin-splitting, $\Delta_{\mathrm{AM}}\left(\mathbf{k}\right)\equiv \varepsilon_{\mathrm{up}}\left(\mathbf{k}\right) - \varepsilon_{\mathrm{dn}}\left(\mathbf{k}\right)$, along the (e) $\Gamma$-R$_1$ and (f)-(h) X$_3$-R$_1$ paths. (i)-(l) Difference in the lifetimes between ``up'' and ``dn'' quasiparticles, $\Delta_\tau\left(\mathbf{k}\right)\equiv \tau_{\mathrm{up}}\left(\mathbf{k}\right)- \tau_{\mathrm{dn}}\left(\mathbf{k}\right) $ and (m)-(p) lifetime $\tau$ as a function of momentum $\mathbf{k}$ along the (m) $\Gamma$-R$_1$ and (n)-(p) $X_3$-R$_1$ paths. \rmf{there is quite a bit of empty space in the left and right sides of this figure. Could we make the panels bigger?} \ina{I modified the size.}}
%     \label{fig:5}
% \end{figure*}

Another interesting correlation effect is displayed in Figs.~\ref{fig:4}(d)-(f), which show that the nominally non-magnetic Se-$p$ states can also develop a sizable AM splitting through strong $d$-$p$ hybridization and in combination with electronic correlation effects. \NS\ is known to experience strong hybridization between Ni-$d$ and S-$p$ orbitals, such that the S-$p$ states have a non-negligible contribution to the Fermi surfaces. ~\cite{kunevs2010metal, moon2015composition} In these figures, since the S-$p$ orbital weight is marked in yellow, the four highest ``up'' and ``dn'' band pairs mainly originate from S-$p$ orbitals (which corresponds to the anti-bonding $\sigma^*$ orbitals of the S-S dimer in the pyrite structure ~\cite{moon2015composition}). 
Along the M$_1$-$\Gamma$ path, these Se-$p$ bands cross and strongly hybridize with the Ni-$e_g$ bands. Consequently, once interactions are included, the Se-$p$ bands also display noticeable modifications. In Fig.~\ref{fig:4}(e), the AM spin splitting is particularly enhanced for the pairs of bands highlighted by blue arrows, which are hybridized between the upper Se-$p$ dominated bands and the lower Ni-$d$ dominated bands. In the DFT+DMFT results in Fig.~\ref{fig:4}(f), these bands are further renormalized by the dynamic correlations, but they still retain a spin splitting of the order of a few hundred meV, which is comparable to that of the Ni-$e_g$ bands. On the other hand, when the S-$p$ bands are not strongly hybridized with Ni-$d$ bands, the AM splitting is negligible, as shown in the uppermost bands in Fig.~\ref{fig:4}(a)-(c). The emergence of such a spin splitting in ligand-derived orbitals demonstrates that altermagnetic properties can extend beyond the correlated $d$ states through hybridization, reinforcing the strongly orbital-dependent nature of the AM splitting.

\begin{figure*}[t]
    \centering
    \includegraphics[width=0.90\linewidth]{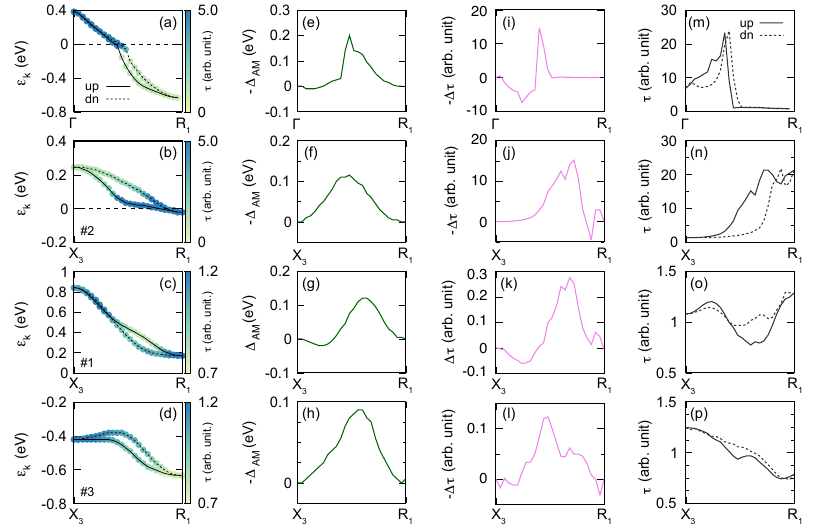}
    \caption{(a)-(d) DFT+DMFT band structures, with the corresponding quasiparticle lifetime marked by colored circles along the (a) $\Gamma$-R$_1$ and (b)-(d) X$_3$-R$_1$ paths. (e)-(h) Altermagnetic spin-splitting, $\Delta_{\mathrm{AM}}\left(\mathbf{k}\right)\equiv \varepsilon_{\mathrm{up}}\left(\mathbf{k}\right) - \varepsilon_{\mathrm{dn}}\left(\mathbf{k}\right)$, along the (e) $\Gamma$-R$_1$ and (f)-(h) X$_3$-R$_1$ paths. (i)-(l) Difference in the lifetimes between ``up'' and ``dn'' quasiparticles, $\Delta_\tau\left(\mathbf{k}\right)\equiv \tau_{\mathrm{up}}\left(\mathbf{k}\right)- \tau_{\mathrm{dn}}\left(\mathbf{k}\right) $ and (m)-(p) lifetime $\tau$ as a function of momentum $\mathbf{k}$ along the (m) $\Gamma$-R$_1$ and (n)-(p) $X_3$-R$_1$ paths.}
    \label{fig:5}
\end{figure*}

\section{\label{sec:r4} 
Correlation effects on the spin-resolved quasiparticle lifetime}

Correlations not only suppress the quasiparticle spectral weight, manifested as a renormalized bandwidth, but also the quasiparticle lifetime, manifested as incoherence. This effect is determined by the frequency dependence of the imaginary part of the self-energy, and therefore can only be captured by approaches that treat the interaction dynamically, like DFT+DMFT. \cite{georges1996dynamical, georges2004strongly, kotliar2006electronic}. In general, the quasiparticle scattering rate nearly vanishes at the Fermi level and increases rapidly with frequency, reflecting phase-space restrictions due to electron-electron scattering. Thus, since the degree of incoherence depends on frequency, and because there is a energy splitting between ``up'' and ``dn'' bands, the two spin bands should display different lifetimes in the altermagnetic state.  

We quantify the lifetime of a band at momentum $\mathbf{k}$ and spin component $\sigma$ by using a Lorentzian fitting to the momentum-resolved spectral function $A_\sigma (\mathbf{k},\omega)$ obtained via DFT+DMFT, which is further described in Appendix~\ref{app:d}. This procedure is equivalent to extracting the inverse scattering rate from the peak width of $A(\mathbf{k},\omega)$, which is directly related to $\mathrm{Im}\Sigma(\omega)$. The results are shown in Fig.~\ref{fig:5}: panels (a)-(d) display selected spin-split bands along high-symmetry directions, colored according to the value of the momentum-dependent lifetime $\tau$. Quite generally, at a given $\mathbf{k}$, the ``up'' and ``dn'' bands exhibit different $\tau$, and the difference is expected to be bigger for larger splitting values. The relationship between these two quantities can be elucidated by comparing Figs.~\ref{fig:5}(e)-(h) and ~\ref{fig:5}(i)-(l), which show, respectively, the momentum-dependent AM spin splitting $\Delta_{\mathrm{AM}}\left(\mathbf{k}\right)\equiv \varepsilon_{\mathrm{up}}\left(\mathbf{k}\right) - \varepsilon_{\mathrm{dn}}\left(\mathbf{k}\right)$ and the difference in lifetime $\Delta_\tau\left(\mathbf{k}\right)\equiv \tau_{\mathrm{up}}\left(\mathbf{k}\right)- \tau_{\mathrm{dn}}\left(\mathbf{k}\right) $. This comparison indicates that the AM splitting does not merely separate the spin bands in energy, but also enforces a contrast in their coherence, effectively locking the energy splitting to the lifetime difference. 

The simplest examples are provided by bands \#1 and \#3 shown in Figs.\ref{fig:5}(b)-(d); their locations within the full band structure manifold are shown in Fig.\ref{fig:4}(c). We find that, when the bands do not cross the Fermi level $E_F$, the momentum dependencies of $\Delta_{\mathrm{AM}}$ and $\Delta_\tau$ strongly correlate with each other.
The general trend that higher binding energy leads to shorter $\tau$ explains the reduced coherence of the spin bands, demonstrating directly how energy separation controls the lifetime difference. Band \#2, on the other hand, shows a different behavior due to the fact that the ``up'' band remains close to $E_F$ over a much wider momentum range than the ``dn'' band. In this momentum range, the ``up'' band has a longer lifetime than the ``dn'' band for being closer to $E_F$, which produces the peak in $\Delta_\tau$ in Fig.~\ref{fig:5}(j).

In contrast to band \#2, both spin bands in Fig.~\ref{fig:5}(a) cross $E_F$ along the $\Gamma$-R$_1$ path at essentially the same momentum value. As shown in Fig.~\ref{fig:5}(i),  $\Delta_\tau$ shows a sharp maximum even though the $k_F$ values of the ``up'' and ``dn'' bands do not differ much. This behavior originates from the strongly electron-hole asymmetric imaginary part of the self-energy. Indeed, as shown in Fig.~\ref{fig:5}(m), the absolute values of both lifetimes suddenly decrease once the bands cross the Fermi value. 

This large $\Delta_{\tau}$ observed around $k_F$ is an effect assisted by Hund's correlations. The importance of correlation effects associated with the Hund's coupling in the paramagnetic phase of \NS\ is well established from previous experimental and theoretical works~\cite{jang2021direct,park2024clean}. In these previous DFT+DMFT calculations in the paramagnetic phase, \NS\ showed a large local moment, an enhanced effective mass, and, as discussed in Appendix~\ref{app:e}, a kink structure at relatively low frequencies in the real part of the self-energy.
In the magnetically ordered state, we find that both majority and minority spin self-energies of \NS\ clearly display similar kink as in the paramagnetic phase  (see Fig.~\ref{fig:s3}(e) and Appendix~\ref{app:e}). Moreover, in the same figure, we also observe the ``inverted slope'' feature, i.e., the sign reversal of the slope of the self-energy near the kink frequency, which is a hallmark of Hund metals \cite{stadler2021differentiating,  kugler2019orbital, kugler2024low}.

The strong particle-hole asymmetry of the majority-spin self-energy results in a large scattering rate in the negative frequency region, as shown in Figs. \ref{fig:s3}(b), \ref{fig:s3}(d), and \ref{fig:s3}(f). In particular, the dramatic enhancement of the scattering rate in the low-energy range from –0.5 eV to 0 eV, which is precisely the energy range corresponding to the bands shown in Fig.~\ref{fig:5}(a), is associated with the presence of a shoulder-like feature in the imaginary part of the self-energy. As we discuss in Appendix~\ref{app:e}, this shoulder-like feature is weakened when the Hund's coupling is decreased, thus establishing the connection between particle-hole asymmetry and Hund's coupling. Because of this effect, even with comparable Ni-$e_g$ orbital weights, the hole-side band suffers a much stronger reduction of $\tau$. This particle-hole asymmetry, combined with the AM splitting, can amplify $\Delta\tau$ near the Fermi level, thus providing a natural mechanism by which AM splitting couples to the anisotropic scattering of each spin band.

\section{\label{sec:conclusion}Conclusion}

In conclusion, \NS\ is a prototypical strongly-correlated altermagnetic material whose pressure or chemical substitution phase diagram hosts an appealing metal–insulator transition within the altermagnetic ordered phase. By combining DFT, DFT+$U$, and DFT+DMFT calculations, we were able to disentangle the effects of static and dynamic correlations on the altermagnetic electronic structure of the metallic phase near the MIT. Static correlations generally results in an enhancement of the local magnetic moment and, consequently, a nearly uniform enhancement of the altermagnetic spin splitting across different bands. In contrast, dynamic correlations described in DFT+DMFT, besides enhancing the local magnetic moment, renormalizes the band mass, resulting in a suppressed bandwidth. 
Because such suppression of the bandwidth (i.e. bandwidth renormalization) depends on both the orbital weight and on energy, the correlation-induced changes in the AM spin splitting varies strongly among different bands. Dynamic correlations also suppress the quasiparticle lifetime of different spin-bands in different ways. Overall, bands with larger spin splitting also have a larger lifetime difference, and this effect can be amplified by the Hund’s correlation–assisted particle–hole asymmetry in the spectrum.

Overall, our work provides a comprehensive survey of how static and dynamic correlations impact different properties of altermagnetic systems, providing guidance for future studies beyond \NS. It also highlights that not only the energies of the spin-up and spin-down quasiparticles can be very difference, but also their spectral weight and their lifetime. This opens a tantalizing scenario in which, for a specific direction, one spin species of quasiparticles could become almost completely incoherent while the other spin species of quasiparticles retain nearly full coherence. This difference between quasiparticle lifetimes could have an important impact on spintronic applications of altermagnets that involve exciting quasiparticles of a particular spin species \cite{Bai2024spintronics,Jungwirth2025spintronics}. More broadly, our work shows the rich landscape of effects that emerge from the interplay between Mott and Hund physics and altermagnetic order.

\begin{acknowledgments}
We thank D. Agterberg for fruitful discussions. R.M.F. was supported by the Air Force Office of Scientific Research under Award No. FA9550-21-1-0423. The Flatiron Institute is a division of the Simons Foundation. T.B. was supported by the NSF CAREER grant DMR-2046020. 
\end{acknowledgments}

\appendix

\section{Crystal Structure of \NS\ at $P$ = 4.3 GPa. \label{app:a}}

Table \ref{tab:1} shows the crystal structure information of \NS\ used in our DFT and DFT+DMFT calculations. These parameters are obtained from the structural relaxation within DFT+DMFT, where the lattice parameter is interpolated from the experimental values \cite{fujii1987structural}. The details of structural relaxation within DFT+DMFT can be found in the reference \cite{park2024clean}.

\begin{table}[h!]
\centering
\caption{Crystal structure of \NS\ at $P$ = 4.3 GPa.}
\label{tab:1}
\begin{tabular}{ccccc}
\toprule
\multicolumn{3}{l}{\textbf{Lattice parameter:} $a = \SI{5.599} {\angstrom}$} & {} & {}\\
\hline
Wyckoff position & Element & x & y & z \\
\hline
$4a$ & Ni1 & 0.000 & 0.000 & 0.000 \\
   & Ni2 & 0.500 & 0.000 & 0.500 \\
   & Ni3 & 0.500 & 0.500 & 0.000 \\
   & Ni4 & 0.000 & 0.500 & 0.500 \\
\hline
$8c$ & S1 & 0.393 & 0.393 & 0.393 \\
   & S2 & 0.607 & 0.607 & 0.607 \\
   & S3 & 0.107 & 0.607 & 0.893 \\
   & S4 & 0.893 & 0.393 & 0.107 \\
   & S5 & 0.893 & 0.107 & 0.607 \\
   & S6 & 0.107 & 0.893 & 0.393 \\
   & S7 & 0.607 & 0.893 & 0.107 \\
   & S8 & 0.393 & 0.107 & 0.893 \\
\bottomrule
\label{tab:1}
\end{tabular}
\end{table}

\section{Computational Details\label{app:b}}
The spin-polarized DFT and DFT+$U$ calculations with full-potential augmented plane wave method were performed by using the WIEN2k code, \cite{blaha2001wien2k,blaha2018WIEN2k, blaha2020wien2k} where the Perdew-Burke-Ernzerhof (PBE) generalized gradient approximation (GGA) was used for the exchange-correlation functional. \cite{perdew1996generalized} For the DFT+$U$ calculations, the around-mean-field (AMF) double-counting scheme was used. A $R_{\mathrm{MT}}K_{\mathrm{max}}$ value of 7 and 12$\times$12$\times$12 $k$-point mesh was used for the electronic self-consistent calculation. 

Charge self-consistent DFT+DMFT calculation was performed as implemented in the DFT + embedded DMFT functional (eDMFTF) code. \cite{haule2010dynamical} For the DMFT calculations, a real harmonics basis was used considering the local axis rotation for each Ni atom and the local octahedral ligand field environment generated by S atoms. For the interaction Hamiltonian, we used the density-density form of the Coulomb interaction with Slater parametrization, $F^0 \equiv U$, $F^2 \equiv \frac{112}{13}J$, and $F^4 \equiv \frac{70}{13}J$ with parameters $U = 8$ eV and $J = 1$ eV. The hybridization window was set from -10 eV to 10 eV. The impurity problem was solved by a continuous-time quantum Monte Carlo (CTQMC) impurity solver. 

The Wannier analysis in Appendix~\ref{app:c} was done using the Quantum Espresso package~\cite{QE} and Wannier90~\cite{pizzi2020wannier90}. The construction of maximally localized Wannier functions (MLWFs) was done using initial projection of Ni-$d_{z^2}$ and Ni-$d_{x^2-y^2}$ atomic orbital basis functions centered at each atomic site and with the corresponding local axis rotation.

\section{DMFT self-energy and Wannier Hamiltonian Analysis \label{app:c}}

\begin{figure*}
    \centering
    \includegraphics[width=0.7 \linewidth]{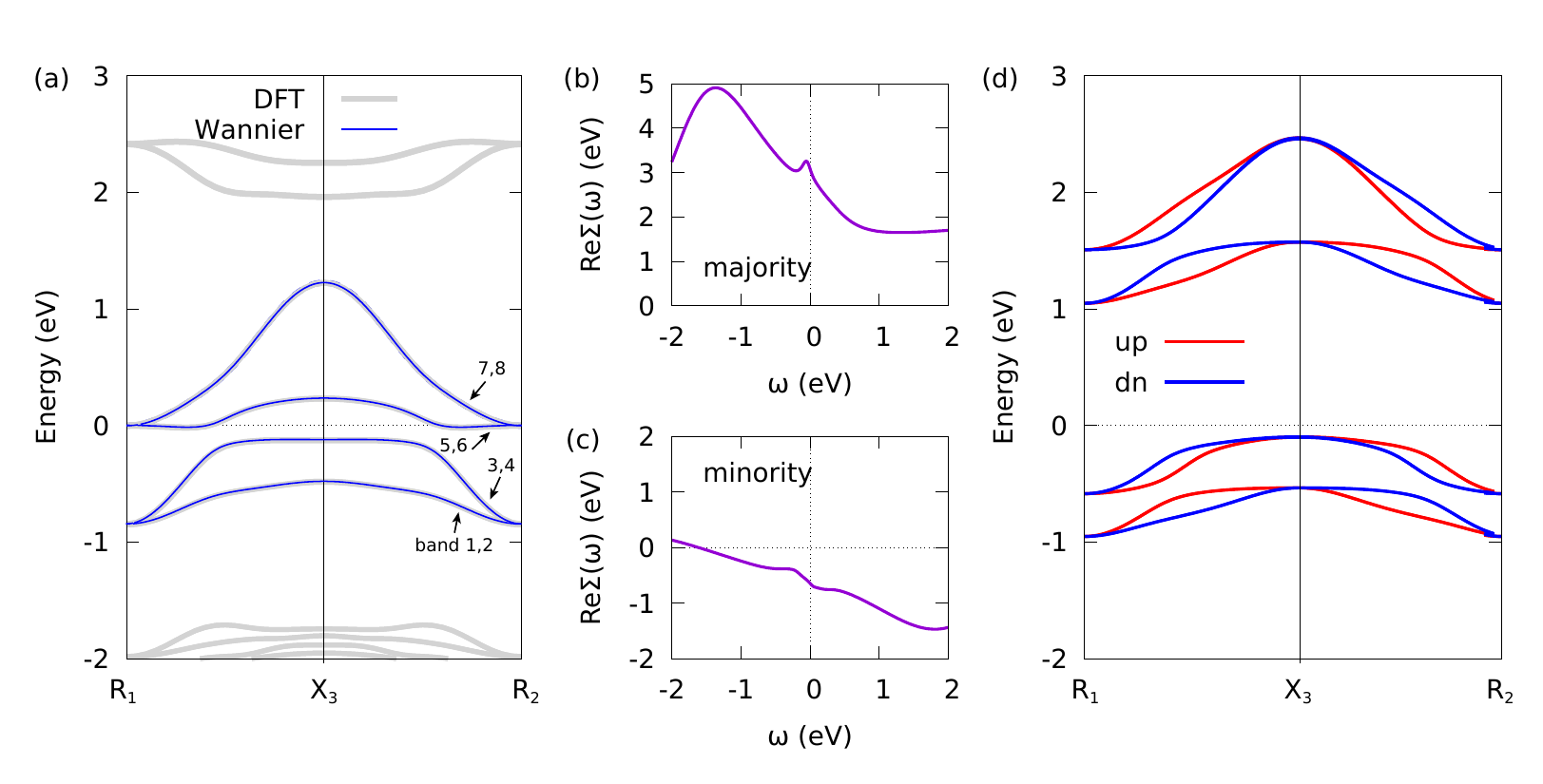}
    \caption{(a) Band structure from DFT calculations (gray thick lines) and the $e_g$ Wannier model (blue lines). Here, $R_1=(-1/2, 1/2, 1/2)$, $X_3=(0, 0, 1/2)$, and $R_2=(1/2, 1/2, 1/2)$. (b)-(c) Real part of the DMFT self-energy Re$\Sigma(\omega)$ for majority and minority spin components. (d) Altermagnetic Wannier band structure. }
    \label{fig:s1}
\end{figure*}

In this Appendix, we provide additional details of the Wannier Hamiltonian and DMFT self-energy analysis of NiS$_2$ that underpins the discussions in the main text. A key finding is that the large spin splitting of the orbital-resolved self-energy at zero frequency, of the order of several electron volts, does not translate as-is into large altermagnetic band splittings. Instead, the spin splitting in the band structure is typically of the order of a few hundred meV. This reduction arises because the effective splitting is determined not by the bare Re$\Sigma(0)$ alone, but by its interplay with the orbital weight of each Bloch state across the Brillouin zone. Because each band state has contributions from different Ni $e_g$ orbital, it experiences distinct projections of the self-energy. This, in turn, renormalizes the spin splitting non-rigidly.  

To demonstrate this, we build a simple $e_g$-only Wannier tight-binding (TB) model and calculate the ``effective'' DMFT band structure. The procedure is as following. As shown in Fig.~\ref{fig:s1}(a), we first extract the $e_g$ Wannier Hamiltonian, where the Hamiltonian for each $\mathbf{k}$ in Wannier orbital basis is denoted as $\varepsilon_{mm'}(\mathbf{k})$ with $m$ and $m'$ denoting orbital indices.  The self-energy is also written in the orbital basis, $\Sigma_{mm'}(\omega)$. For the sake of simplicity, although the fully charge self-consistent DFT+DMFT calculation yields the full 5x5 (or 10x10 with spin) Ni $d$-orbital self-energy, in the present Wannier analysis, we retain only its $e_g$ block. The real part of the self-energy for majority and minority spin-component is shown, for example, in Fig.~\ref{fig:s1}(b) and ~\ref{fig:s1}(c).

\begin{figure*}
    \centering
    \includegraphics[width=0.7\linewidth]{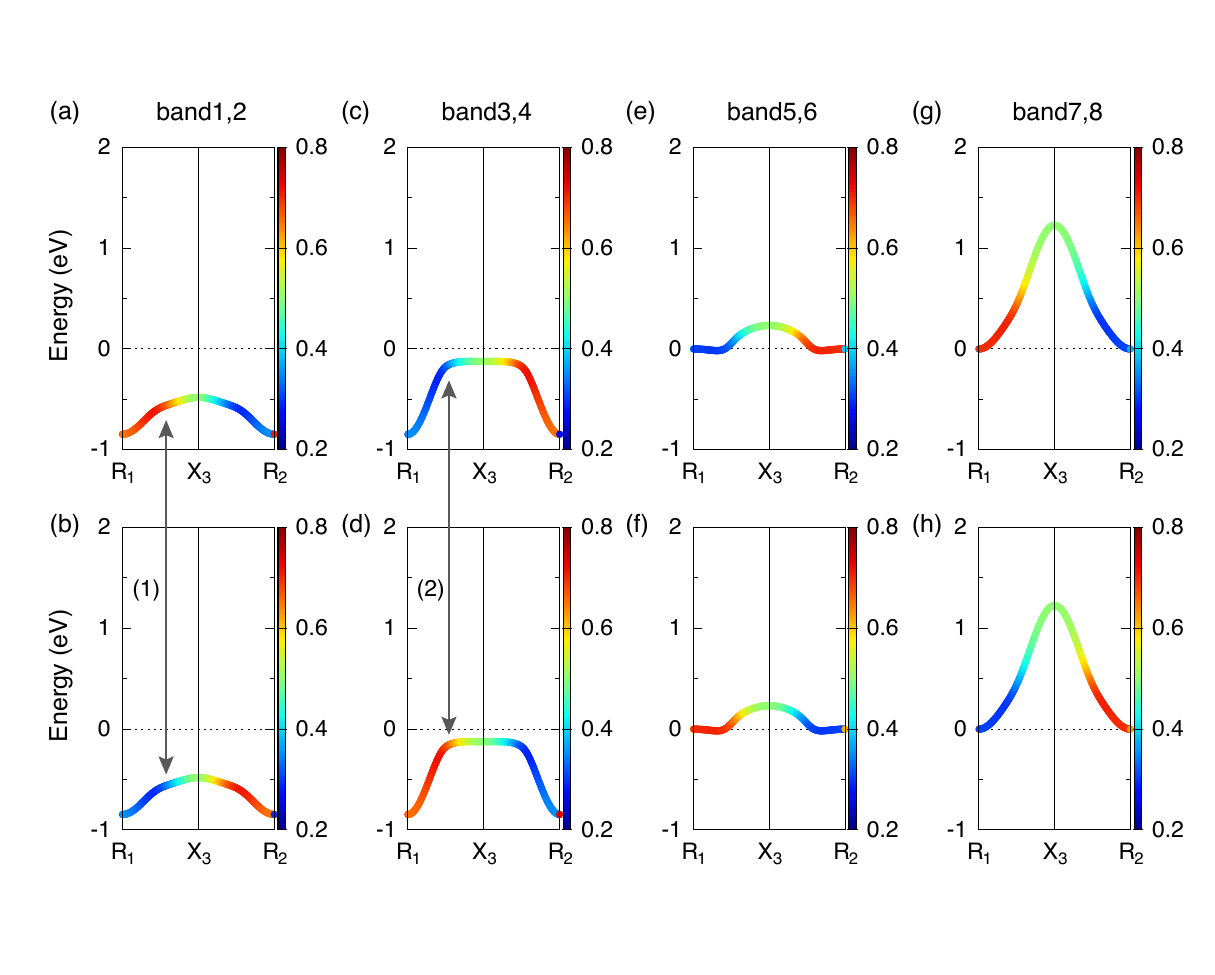}
    \caption{Band structure projected on the sublattices of the Ni atoms (see Fig. \ref{fig:1}(d)-(e)) in the paramagnetic phase. Top row shows the $e_g$-orbital weight of Wannier orbitals centered at the Ni1 and Ni2 sites, whereas the bottom row shows the orbital weights for Ni3 and Ni4 sites. Bands 1-8 are marked in Fig.\ref{fig:s1}(a). }
    \label{fig:s2}
\end{figure*}

From $\varepsilon_{mm'}(\mathbf{k})$ and $\Sigma_{mm'}(\omega)$, the Green's function can be built as (now with spin indices for clarity): 
\begin{equation}
    G_{ mm', \sigma}^{-1}(\mathbf{k},\omega) =  \omega + \mu - \varepsilon_{mm'}(\mathbf{k}) - \Sigma_{mm'}^\sigma(\omega),
    \label{eqn:green's function}
\end{equation}

where the self-energy for each spin $\sigma$ is given by:
\begin{equation}
\Sigma^{\sigma}(\omega)=
\begin{pmatrix}
\Sigma^{\sigma}_{1}(\omega) & 0 & 0 & 0\\
0 & \Sigma^{\sigma}_{2}(\omega) & 0 & 0\\
0 & 0 & \Sigma^{\sigma}_{3}(\omega) & 0\\
0 & 0 & 0 & \Sigma^{\sigma}_{4}(\omega)
\end{pmatrix}.
\label{eqn:self-energy}
\end{equation}

In this equation, each block $\Sigma_i^\sigma(\omega)$ represents the self-energy of the $i$-th Ni atom,
\begin{equation}
\Sigma^{\sigma}_{i}(\omega) \equiv
\left.
\Sigma^{\sigma}_{i,mm'}(\omega) \delta_{mm'}
\right|_{m,m'\in e_g},   % \in\mathbb{C}^{2\times2},
\label{eqn:self-energy2}
\end{equation}
with only the $e_g$ components taken into account. As explained in Sec.\ref{app:b}, we diagonalized the self-energy block matrices by appropriately rotating the local axis of each impurity atoms.
These blocks satisfy the relations imposed by the altermagnetic ordering, defined as:

\begin{equation}
\begin{aligned}
\Sigma^{\uparrow}_{i, mm}(\omega) &=
\begin{cases}
\Sigma_{\mathrm{maj}}(\omega), & i=1,2\\
\Sigma_{\mathrm{min}}(\omega), & i=3,4
\end{cases}
\\
\Sigma^{\downarrow}_{i, mm}(\omega) &=
\begin{cases}
\Sigma_{\mathrm{min}}(\omega), & i=1,2\\
\Sigma_{\mathrm{maj}}(\omega), & i=3,4
\end{cases}
\\
& \quad (m \in e_g).
\end{aligned}
\label{eqn:self-energy3}
\end{equation}

From Eqs.~\ref{eqn:self-energy}--\ref{eqn:self-energy3}, one can clearly see how the altermagnetic ordering of the four Ni atoms in the unit cell is imposed within the DFT+DMFT through the structure of the self-energy. The altermagnetic ordering is imposed by relating the spin- and site-resolved self-energies according to Eq.~\ref{eqn:self-energy3},  i.e.

\begin{equation}
\begin{aligned}
    \Sigma^{\uparrow(\downarrow)}_{1, mm}(\omega) &= \Sigma^{\uparrow(\downarrow)}_{2, mm}(\omega) = \Sigma^{\downarrow(\uparrow)}_{3, mm}(\omega) = \Sigma^{\downarrow(\uparrow)}_{4, mm}(\omega).
\end{aligned}
\end{equation}

The spin-up self-energy of the Ni1 and Ni2 sites is identical to the spin-down self-energy of the Ni3 and Ni4 sites, identified as the majority spin self-energy, $\Sigma_{\mathrm{maj}}(\omega)$. Conversely, the spin-down self-energy of Ni1 and Ni2 sites is identical to the spin-up self-energy of Ni3 and Ni4 sites, corresponding to the minority spin self-energy, $\Sigma_{\mathrm{min}}(\omega)$. This imposed structure of the self-energy in DFT+DMFT calculation captures the defining feature of altermagnetism, the vanishing net magnetization despite a spin-dependent band structure. 

% The resulting electronic structure by embedding this self-energy into the Green's function (Eq.~\ref{eqn:green's function}) shows clear spin split ``up'' and ``dn'' bands shown in Fig.~\ref{fig:s1}(d).}

Back to the procedure, after building the Green's function, diagonalization of the matrix in Eq.~\ref{eqn:green's function} gives eigenvalues $\lambda_{\nu}(\mathbf{k},\omega)$ at each $\omega$ and $\mathbf{k}$ ($\nu$ is band index), and for each $\mathbf{k}$, poles (roots) can be calculated. This is equivalent to solving Eq.~\ref{eqn:3} in the main text. The resulting pole frequencies $\omega_{\mathbf{k},\nu}$ become the final effective altermagnetic DMFT band structure, which is shown in Fig.~\ref{fig:s1}(d).

This analysis can explain the resulting altermagnetic spin structure, including the few hundreds meV magnitude of the spin splitting and the momentum-dependent sign reversal of the spin splitting, based on the orbital projection of the bands. 
From the orbital weights of band $\nu$ at a $\mathbf{k}\textbf{}$ point, $\langle\chi_{m}|\psi_{\mathbf{k}\nu}\rangle$, one can extract the self-energy at each $\mathbf{k}$ for band $\nu$:
\begin{equation}
    \Sigma(\mathbf{k}, \omega)_{\nu\nu'} = \sum_{mm'} {\langle \psi_{\mathbf{k}\nu}| \chi_m\rangle \Sigma_{mm'}(\omega)\langle\chi_{m'}|\psi_{\mathbf{k}\nu'}\rangle }. 
\end{equation}

% \begin{figure*}
%     \centering
%     \includegraphics[width=0.7\linewidth]{Figure S2.pdf}
%     \caption{Band structure projected on the sublattices of the Ni atoms (see Fig. \ref{fig:1}(d)-(e). Top row shows the $e_g$-orbital weight of Wannier orbitals centered at the Ni1 and Ni2 sites, whereas the bottom row shows the orbital weights for Ni3 and Ni4 sites. \added{Bands 1-8 are marked in Fig.\ref{fig:s1}(a).} \rmf{Could we please label these bands 1,2,3,4 etc in Fig. S1? Also could we please label the panels as (a),(b), (c), etc?} \ina{Yes, I added guiding text in Fig.S1(a) and captions here. I also explanation in the figure caption here.}}
%     \label{fig:s2}
% \end{figure*}

Note that the momentum dependence of the self-energy comes from the orbital weights. In Fig.~\ref{fig:s2}, we plot the atom-projected weight of each band. The top row shows the total weight from the $e_g$ orbitals of atoms Ni1 and Ni2 (up spins in Fig.~\ref{fig:1}(d)-(e)), whereas the bottom row has the $e_g$-orbital weight of Ni3 and Ni4 (down spins in Fig. \ref{fig:1}(d)-(e)). For bands 1,2 in Fig.\ref{fig:s2}(a)-(b), the lowest bands, the contribution from Ni1 and Ni2 is more dominant than that from Ni3 and Ni4 along the $R_1$–$X_3$ path [see arrow (1)]. Conversely, along the $X_3$-$R_2$ path, the opposite happens, and Ni3(4) is more dominant than Ni1(2). In the altermagnetic phase, the ``up'' spin band (bottom red curve in Fig.~\ref{fig:s1}(d)) is obtained from this orbital composition and from $\Sigma_{mm'}^{\mathrm{up}}$. For the ``dn'' spin band, while the orbital composition does not change, the self-energy changes to $\Sigma_{mm'}^{\mathrm{dn}}$. In contrast, for bands 3,4 in Fig.\ref{fig:s2}(c)-(d), it is Ni3(4) that gives the dominant contribution along the $R_1$-$X_3$ path, as highlighted by arrow (2). Consequently, for bands 3,4, it is the ``dn'' band that has a higher energy than the ``up'' band along the $R_1$-$X_3$ path, as shown in Fig.~\ref{fig:s1}(d).

\begin{figure*}
    \centering
    \includegraphics[width=0.8\linewidth]{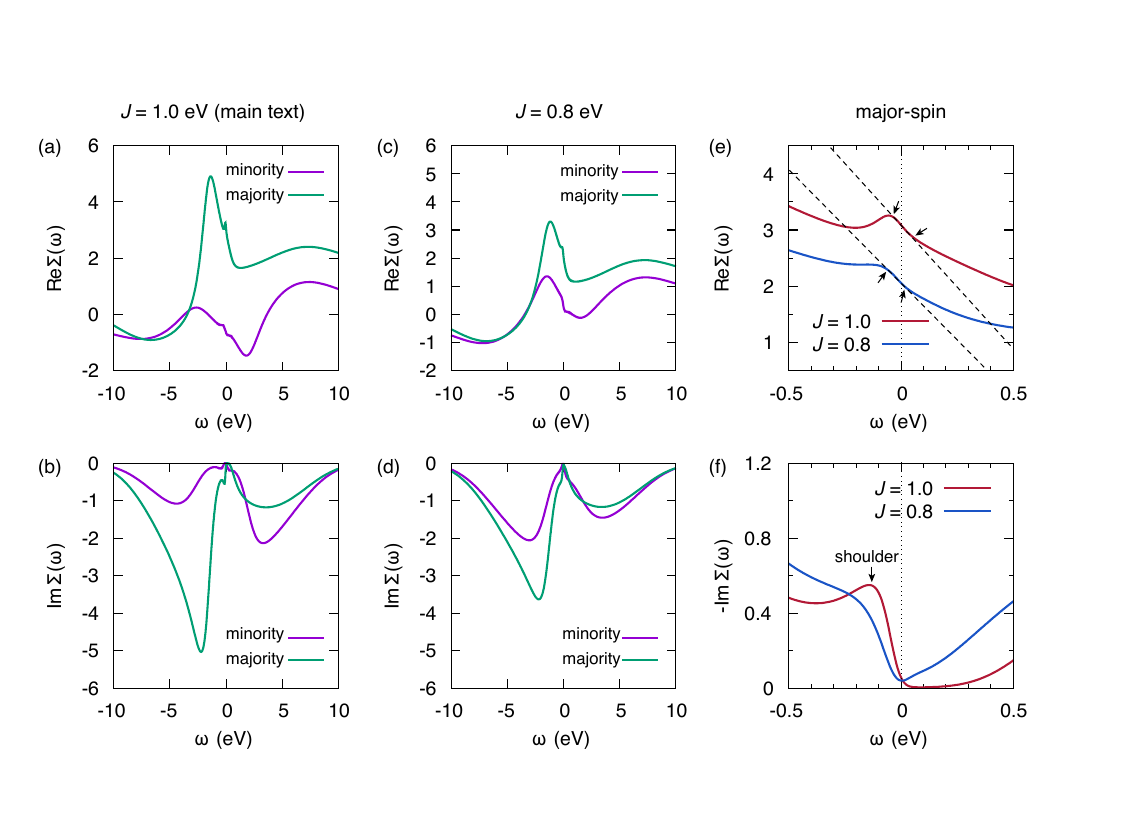}
    \caption{DFT+DMFT self-energy for different $J$ values. (a)-(b) Real (a) and imaginary (b) parts of the DMFT self-energy for majority and minority spin components for $J=1.0$ eV. (c)-(d) Real (c) and imaginary (d) parts of the DMFT self-energy for majority and minority spin components for $J=0.8$ eV. (e)-(f) Real (e) and imaginary (f) parts of the DMFT self-energy of the majority spin component over a narrower frequency range. In (e), the dashed line denotes the low-frequency linear behavior that characterizes the quasiparticles and the arrows mark the ``kink'' where the self-energy starts to deviate from the linear behavior. In (f), the arrow marks the ``shoulder'' feature. } 
    \label{fig:s3}
\end{figure*}

This orbital-projection analysis shows that different Ni sublattices contribute asymmetrically to each band, so that after projecting the self-energy onto the band basis, the spin-splitting can appear with opposite signs. As a result, certain band pairs display an inverted order of up and down spin components along particular momentum paths, which, in the case of \NS, reflects the $d$-wave symmetry expected from the group theory analysis. This explains why the sign of the altermagnetic splitting, rather than being fixed globally, can reverse depending on the orbital character and path in momentum space.

% \begin{figure*}
%     \centering
%     \includegraphics[width=0.7\linewidth]{Figure S3.pdf}
%     \caption{DFT+DMFT self-energy for different $J$ values. (a)-(b) Real (a) and imaginary (b) parts of the DMFT self-energy for majority and minority spin components for $J=1.0$ eV \rmf{could you please change major to majority or maj and similarly for minor in the figure?}. \ina{I changed that.} (c)-(d) Real (c) and imaginary (d) parts of the DMFT self-energy for majority and minority spin components for $J=0.8$ eV. (e)-(f) Real (e) and imaginary (f) parts of the DMFT self-energy of the majority spin component over a narrower frequency range. In (e), the dashed line denotes the low-frequency linear behavior that characterizes the quasiparticles and the arrows mark the ``kink'' where the self-energy starts to deviate from the linear behavior. In (f), the arrow marks the ``shoulder'' feature. \rmf{this figure is in a separate page after all appendices. I tried to move it around, but the compiler keeps putting it in this weird place. Could we force it to be on the previous page?}} 
%     \label{fig:s3}
% \end{figure*}

\section{Lorentzian fitting for extraction of DMFT effective bands and quasiparticle lifetimes \label{app:d}}

The quasiparticle DMFT band structure and the $\mathbf{k}$-dependent lifetime discussed in the main text were extracted from the DMFT spectral function $A(\mathbf{k},\omega)$ using a two-step procedure described in this section.

\textbf{(i) Quasiparticle band centers}

We first constructed a ``coherent'' reference spectral function $A_{0}(\mathbf{k}, \omega)$ by setting Im$\Sigma(\omega) \sim 0$ and only using the value of Re$\Sigma(\omega)$. Then, for each momentum $\mathbf{k}$, we obtained $\mathbf{k}$ cuts of the DMFT spectral function, $A_{0}(\mathbf{k},\omega)$ and extracted the peak positions $\{\omega_{0,i}\}_\mathbf{k}$ using peak finding implemented in the SciPy Python library, where $i$ is the peak index. The peaks of this spectrum, $\{\omega_{0,i}\}_\mathbf{k}$,  were then used to define the band centers shown in Fig.~\ref{fig:4}. This approach provides a stable backbone for the quasiparticle band dispersion by removing the effect of the imaginary part of the self-energy, which controls the peak broadening. We note that, in general, one may use $\omega$ cuts rather than $\mathbf{k}$ cuts, but in the case of \NS\ since the bands are relatively  flat due to the strong correlations renormalization, peak finding using $\omega$-cuts is in practice challenging. Therefore, we instead used $\mathbf{k}$ cuts, which enable us to find clear peak positions.

\textbf{(ii) Momentum-dependent lifetime}

Using the above peak as fixed centers, the full interacting spectral function $A(\mathbf{k},\omega)$ for each momentum $k$ was decomposed into a sum of Lorentzian functions,
\begin{equation}
    A(\mathbf{k},\omega) \approx \sum_i { \frac{A_{i,\mathbf{k}}}{\pi} \frac{\gamma_{i,\mathbf{k}}}{\left( \omega-\omega_{0,i,\mathbf{k}}\right)^2 + \gamma_{i,\mathbf{k}}^2}}.
\end{equation}

As mentioned, the peak centers $\omega_{0,i,k}$ were fixed from step (i), while the  $\gamma_{i,\mathbf{k}}$ and amplitudes $A_{i,\mathbf{k}}$ were optimized. The momentum-dependent scattering time is then defined as $\tau_{i,\mathbf{k}} \equiv \left[ 2\gamma_{i,\mathbf{k}} \right]^{-1}$. This approach using fixed centers avoids spurious peak shifts during the Lorentzian fitting and ensures that the extracted widths faithfully capture the effect of Im$\Sigma(\omega)$. We confirmed that the sum of all Lorentzian functions reproduced the original spectral function reasonably well.

\section{Hund's correlation effects and particle-hole asymmetry in the DMFT self-energy \label{app:e}}

As discussed in the main text, paramagnetic \NS\ is known to exhibit strong Hund’s correlation effects due to its multi-orbital nature.~\cite{georges2013strong} A representative signature is the so-called kink structure in Re$\Sigma(\omega)$. Such kink structures were also reported in ARPES studies of Se-doped \NS\ and identified to arise from Hund’s correlations via comparison with DFT+DMFT calculation.~\cite{jang2021direct} In stoichiometric \NS, the same kink structures were also found in DFT+DMFT calculations near the pressure-induced MIT.~\cite{park2024clean}

In altermagnetic \NS\ near the MIT, we find that Re$\Sigma(\omega)$ still shows a clear kink, similar to the paramagnetic case ~\cite{park2024clean}. This is shown in Fig.~\ref{fig:s3}(a), which displays the real part of the majority and minority self energies over an extended energy range. To further investigate this kink, Fig.~\ref{fig:s3}(e) shows that the majority spin self-energy for $J=1.0$ eV, which is the value used in the main text, follows a linear-in-frequency behavior in a very low-energy range near the Fermi level (marked with dashed lines). Its deviation from linear behavior (marked by arrows) away from this low-energy range is what gives rise to the kink. 

The imaginary part of the majority and minority self-energies are shown in Fig.~\ref{fig:s3}(b). The pronounced particle–hole asymmetry of the scattering rate mainly originates from the large (in absolute value) majority-spin Im$\Sigma(\omega)$ at negative frequencies. This can be understood as the majority spin experiencing more scattering events in the negative (filled) frequency region as compared to the positive frequency region. This asymmetry is also accompanied by a different degree of correlation between the two spin channels. Indeed, the majority spin has effective mass $m^{\mathrm{maj}}_{\mathrm{eff}} \sim 6.25$ and lifetime $\Gamma/k_BT \equiv -Z\mathrm{Im}\Sigma(0^+)= 2.08$, whereas the minority spin has $m^{\mathrm{min}}_{\mathrm{eff}} \sim 2.55$ and $\Gamma/k_BT = 0.70$. Finally, the kink feature in the spin-majority Re$\Sigma(\omega)$ is also reflected in the spin-majority Im$\Sigma(\omega)$ as a ``shoulder'' feature via the Kramers–Kronig relations \cite{stadler2015dynamical, kugler2019orbital}, marked with an arrow in Fig.~\ref{fig:s3}(f). 

When the Hund's coupling $J$ is decreased to 0.8 eV, the size of the local moment and the magnitude of the exchange splitting decrease, see Fig.~\ref{fig:s3}(c). Moreover, as shown in Fig.~\ref{fig:s3}(d),  Im$\Sigma(\omega)$ shows a much less pronounced particle-hole asymmetry, suggesting that scattering events on the negative (filled) frequency region are suppressed. Figs.~\ref{fig:s3}(e) and ~\ref{fig:s3}(f) further show that smaller $J$ values produce much less prominent kink and shoulder features in the low-energy regime of the real and imaginary self-energies, respectively. Therefore, for low-energy bands, such as the band in Fig.~\ref{fig:5}(a), the prominent particle-hole asymmetry arises not only from the reduced positive-frequency scattering due to band filling, but also from the Hund’s-induced enhancement of negative-frequency scattering manifested in the kink and shoulder features in the spin-majority self-energy.

In conclusion, these results indicate that Hund’s coupling plays two essential roles: (i) controlling the exchange splitting and overall particle-hole asymmetry of scattering, and (ii) producing low-energy kink and shoulder structures in the self-energy that further enhance the asymmetry at low energies.

\clearpage
\bibliography{reference}% Produces the bibliography via BibTeX.

\end{document}